\documentclass[11pt,twoside]{article}
 
\usepackage{asp2006}
\usepackage{epsf}
\usepackage{psfig}
\usepackage{lscape}
 
\begin{document}

\title{Tidal Dwarf Galaxies, Accretion Tails,
and `Beads on a String' in the `Spirals, Bridges, and Tails'
Interacting Galaxy Survey}
\author{Beverly J. Smith$^1$,
Mark L. Giroux$^1$,
Curtis Struck$^2$,
Mark Hancock$^{3}$, and
Sabrina Hurlock$^1$}
\affil{$^1$East Tennessee
State University, 
$^2$Iowa State University, 
$^3$University of California Riverside}

\begin{abstract} 

We have used the GALEX ultraviolet telescope to study stellar populations and 
star formation morphology in a well-defined sample of more than
three dozen nearby 
optically-selected 
pre-merger 
interacting galaxy pairs. 
We have combined the GALEX NUV and FUV 
images with broadband optical maps from the Sloan Digitized Sky
Survey to investigate
the
ages and extinctions of the tidal features
and the disks.  
We have identified a few new
candidate tidal dwarf galaxies in this sample, as well as
other interesting morphologies
such as accretion tails, `beads on a string', and `hinge clumps'.
In only a few cases are strong tidal features seen in HI maps
but not in GALEX.

\end{abstract}



\section{Introduction}

Tidal disturbances 
have played an 
important role in reshaping galaxies and triggering star formation
over 
cosmic time. 
This is confirmed by 
H$\alpha$,
far-infrared, 
and mid-infrared
studies showing that the 
mass-normalized star formation rates
of pre-merger
optically-selected interacting
galaxies 
are enhanced by a factor of two on average compared to normal spirals
\citep{bushouse87, kennicutt87, bushouse88,
smith07}.

With the advent of the Galaxy Evolution Explorer (GALEX), a new window
on star formation in galaxies is now available.
The addition
of UV helps to break the age$-$extinction degeneracy in population
synthesis modeling (e.g., \citealp{smith08}).
Furthermore,
since the UV traces 
somewhat older and lower mass stars ($\le$400 Myrs; O to early-B stars) than 
H$\alpha$ ($\le$10 Myrs; early- to mid-O stars), it provides a measure of star formation
over a longer timescale than H$\alpha$ studies.
GALEX imaging 
has shown that some tidal features in interacting galaxies
are quite bright in the UV (e.g., \citealp{neff05}).
In some cases, 
tidal features previously thought to be purely gaseous 
have been detected by GALEX (e.g., \citealp{hancock07}).
In other systems, GALEX images
have been used to identify new tidal features 
(e.g., \citealp{boselli05}).

To address these issues, we have used the 
GALEX telescope to image 
more than three dozen strongly 
interacting galaxies in the UV (the `Spirals, Bridges,
and Tails' (SB\&T) sample).
These galaxies were selected from 
the Arp (1966) Atlas 
using the following criteria:
1) They are relatively isolated binary systems; we 
eliminated merger remnants, close triples, and multiple
systems in which the galaxies have similar
optical brightnesses. 
2) They are tidally disturbed.
3)
They have
radial velocities less than 10,350 km/s.
4) Their total angular size is $>$ 3$'$, to allow for 
good resolution with
GALEX.   

Each galaxy was imaged for $\ge$1500 seconds in the 
FUV and 
NUV broadband filters of GALEX, which 
have effective bandpasses of 
1350 $-$ 1705\AA~ and
1750 $-$ 2800\AA, 
respectively.
Some of the galaxies that fit our selection criteria were previously
observed by guaranteed time projects.  For these galaxies,
we used the archival GALEX images.
The circular GALEX field of view has a diameter of 1.2 degrees.
The pixel size is 1\farcs5, and the spatial resolution $\sim$ 5$''$.
About 2/3rds of our galaxies
have broadband
optical images available from the Sloan Digitized Sky Survey (SDSS),
while 3/4 have 
broadband Spitzer infrared
images available \citep{smith07}.
About half have 
published
21 cm HI maps.

\section{Morphologies}

The SB\&T galaxies have a large range of collisional morphologies, 
including 
M51-like systems, wide pairs with long tails and/or bridges, 
wide pairs with short tails, close pairs with 
long tails, and close pairs with short tails.
In the current paper, we discuss unusual tidal morphologies
in a subset of the galaxies.   In \citet{giroux10},
we 
present an Atlas of UV images of additional SB\&T galaxies.
The full survey is described in detail
in \citet{smith10}.
For four of the galaxies in the SB\&T sample,
we have already published
the GALEX images as part of 
detailed studies of the individual
galaxies, and compared with numerical simulations of the interaction
\citep{hancock07, hancock09, hancock10, smith08, peterson09, peterson10}.

There is a large variety of star formation
morphologies within the tidal features in this sample.
In many cases, the tidal features are quite bright in the UV.
This is illustrated by Arp 72 (Figure 1a),
whose eastern tail is very prominent in the GALEX images,
and has very blue UV/optical colors.
Arp 72 is also a good example of 
the so-called `beads on a string' morphology, in which 
regularly-spaced clumps of star formation are seen along
spiral arms and tidal features.
These clumps are generally spaced about 1 kpc apart, 
the characteristic scale for gravitational collapse
of molecular clouds \citep{elmegreen96}.
Such beads are seen in 
many other systems in our sample,
including the northern tail of 
the western galaxy in Arp 65 (Figure 1b), Arp 82 \citep{hancock07},
and Arp 285 \citep{smith08}.

\begin{figure}
\plottwo{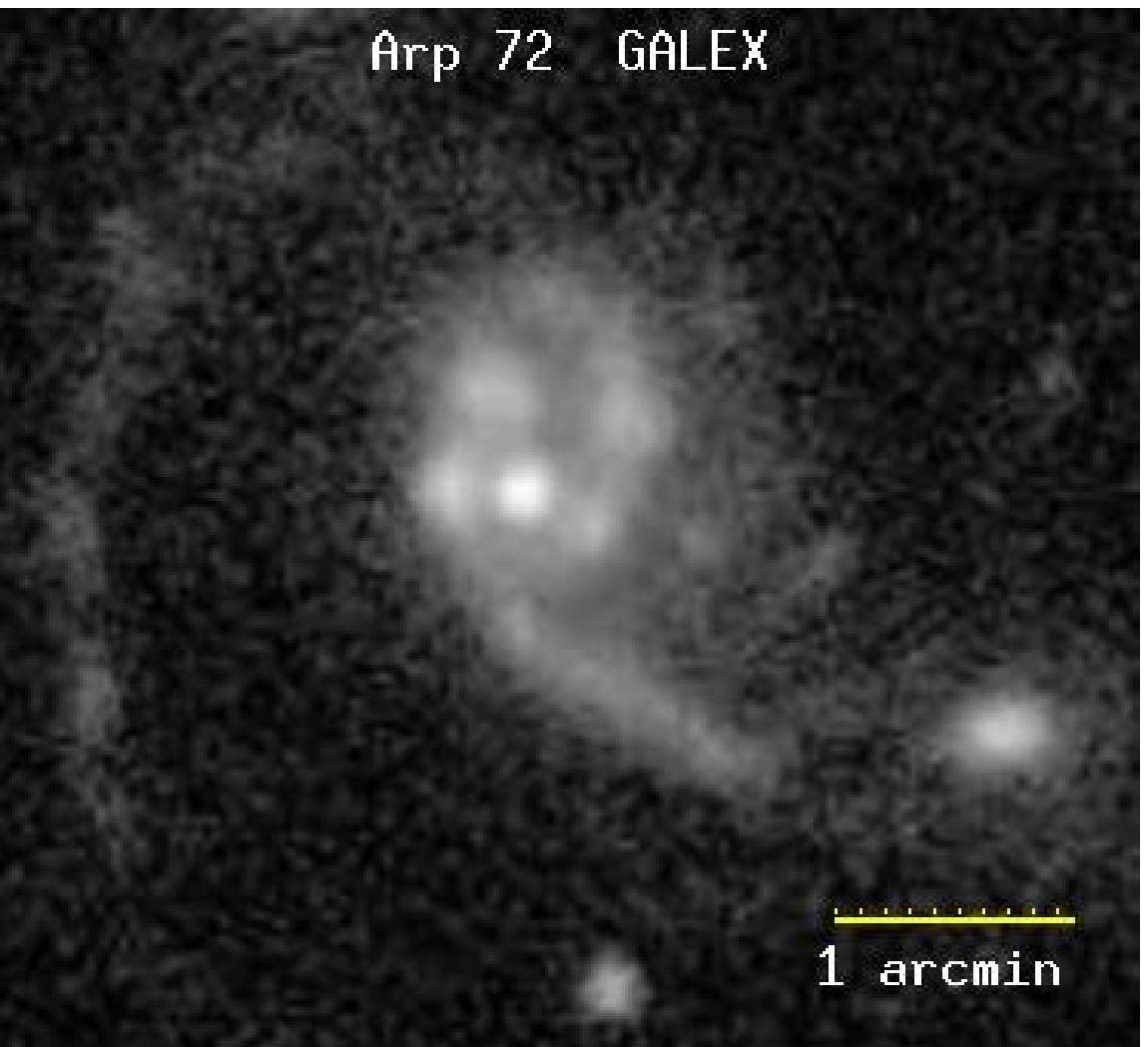}{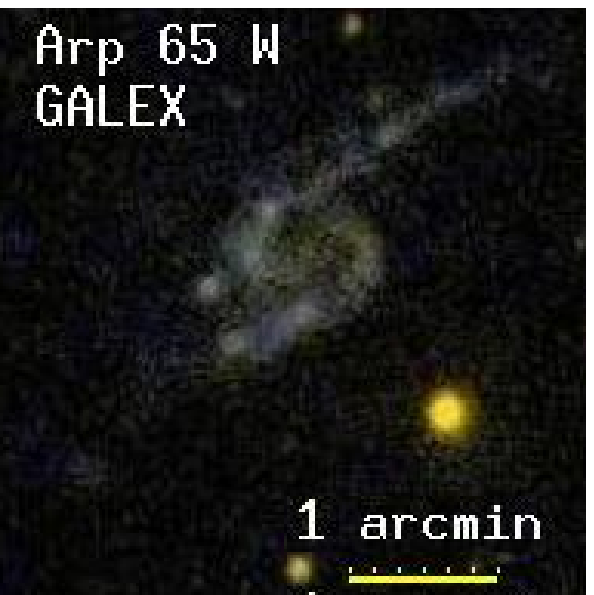}
\caption{
  \small 
GALEX images of Arp 72 (left) and the western galaxy of Arp 65 (right).
North is up and east to the left in all figures in this paper.
}
\end{figure}

In a few systems, we see luminous star forming regions
at the base of a tidal feature.  We call these features
`hinge clumps' \citep{hancock09}.   These lie near the
intersection of the spiral density wave in the inner disk
and the material wave in the tail.   These may form
when dense material in the inner disk gets pulled out
into a tail.   This lowers the shear, which may allow
more massive clouds to gravitationally collapse.
Hinge clumps are visible at the eastern end of the 
Arp 72 bridge (Figure 1a) and the base of the northern tail
of Arp 65 (Figure 1b).  Hinge clumps are also seen in
Arp
82 \citep{hancock07} and Arp 305 \citep{hancock09}.

Our sample also includes some candidate `tidal dwarf galaxies' (TDGs),
massive concentrations of
young stars near the tips of tidal features.
The prototypical TDGs
in Arp 244 and Arp 245
\citep{mirabel92, duc00}
are included in the SB\&T sample,
along with the bridge TDG in Arp 305 
\citep{hancock09, hancock10}.
Another possible TDG is seen in 
Arp 202 (Figure 2), an interaction
between an edge-on disk galaxy and a smaller
irregularly-shaped galaxy to the south.
A long clumpy tail is visible 
to the west of the southern galaxy.   
The tip of this tail is particularly prominent in
the GALEX images, and has very blue UV/optical colors.
Our optical spectrum shows that
this clump
is at the same redshift as Arp 202.
This source
was not detected
in our Spitzer 8 $\mu$m map \citep{smith07}
or in our SARA H$\alpha$ map, 
suggesting that it is in a post-starburst stage.

\begin{figure}
\plottwo{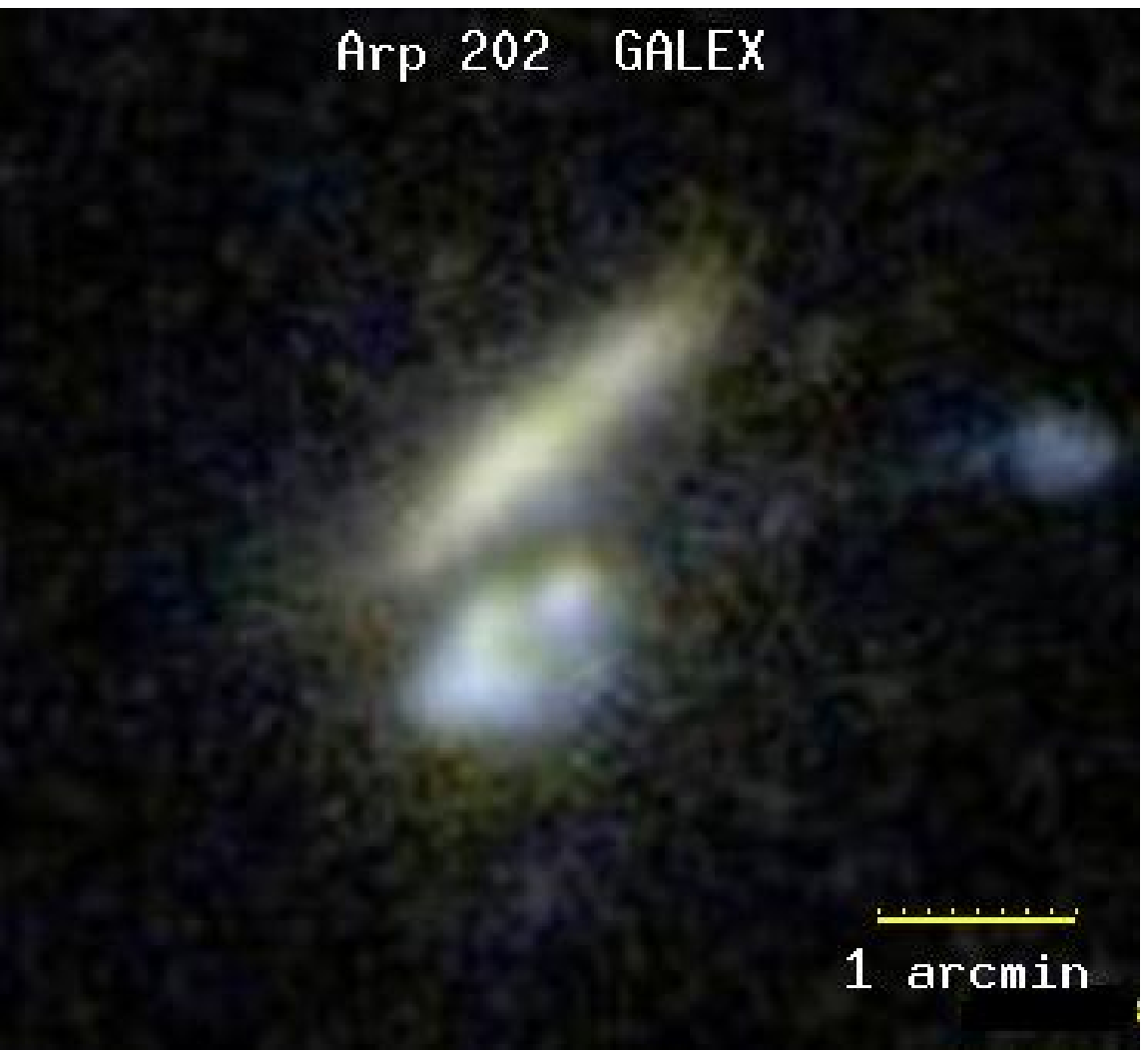}{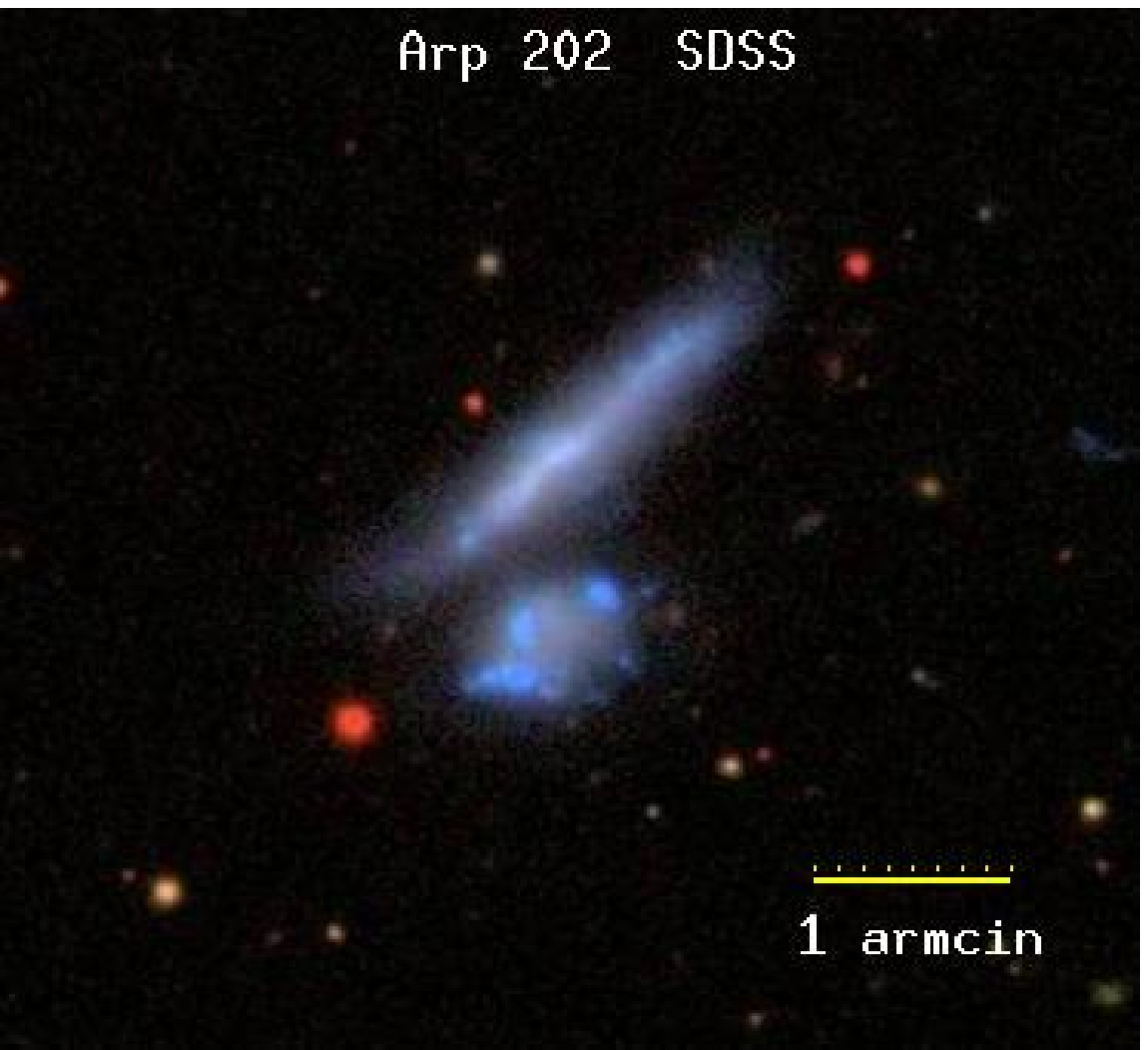}
\caption{
  \small 
The GALEX (left) and SDSS (right) images of Arp 202.
}
\end{figure}

Another SB\&T system that may have TDGs is Arp 181 (Figure 3).
A clump is visible 
near the end of the western tail 
in the GALEX and SDSS images, with
very blue optical/UV colors.
However, no optical spectrum
is available,
thus it is unclear whether it is at the same redshift as
Arp 181.
Further west,
another galaxy is visible, without 
any obvious link to the
tail.
Our optical spectrum 
shows that it 
is at the same redshift as Arp 181.  
In the SDSS image it
looks like a spiral galaxy or a disturbed disk with short tidal tails.  
It is extremely blue in NUV $-$ g, and is detected at 8 $\mu$m \citep{smith07}.
This may be either a pre-existing dwarf
galaxy or a recently detached TDG.

The SB\&T sample also contains numerous examples of accretion
from one galaxy to another.   One of the best-studied
examples is the northern tail of Arp 285,
which was likely produced from material accreted from the 
southern galaxy
\citep{toomre72, smith08}.
According to our numerical simulations, the material in this
tail fell into
the gravitational potential of the northern galaxy, overshot
that potential, and is now gravitationally collapsing and forming
stars \citep{smith08}.
We call such features `accretion tails', to distinguish
them from classical tidal features.  The inner western
tail of Arp 284 was likely produced by the same 
process \citep{struck03}.   
Another system which may have an accretion tail is Arp 105
(Figure 4).  
The spiral in this system has a long tail
extending to the north, previously classified
as a TDG \citep{duc97}.
The spiral is connected 
by a bridge
to an elliptical 
galaxy to the south.
South of the elliptical 
is a bright star formation knot
\citep{stockton72}.
Both the northern TDG candidate and the southern
knot of star formation are luminous in HI maps
\citep{duc97}.
In the GALEX images, 
the spiral and the northern TDG are quite bright in the UV,
but the highest UV surface brightness is found in
the knot of star formation south of the elliptical.
We suggest, based on analogy to Arp 285 \citep{smith08} and 
proximity to the elliptical, that the southern
star forming region in Arp 105 is an accretion tail, 
rather than
simply a classical tidal tail coincidently seen in projection
behind the elliptical.

Another system with possible mass transfer is 
Arp 269 (Figure 5a), an unequal-mass pair of galaxies with
a bridge.   In both the GALEX and SDSS images,
an off-center group of blue star forming regions is visible
in the smaller galaxy NGC 4485, as well as
along the bridge.   
As noted by \citet{elmegreen98}, several of these knots of
star formation lie in a tail-like structure 
southwest of NGC 4485.
\citet{clemens00} suggest that
NGC 4485 passed through
the 
disk of the larger galaxy NGC 4490, and ram pressure 
caused an offset in the location of the interstellar
gas in NGC 4485, and thus the observed offset in star formation.
We suggest an alternative possibility, 
that
the star formation
was triggered by
gas flow
from NGC 4490 along the bridge.
Thus this may be an example of accretion from one galaxy to another.

Our sample contains only a few tidal features 
with high HI column densities 
that are not detected in
our GALEX maps or published optical maps.
One of these systems is Arp 269.
In the \citet{clemens98a} 
HI maps, two large plume-like features
extend 10$'$ ($\sim$20 kpc)
to the north and south of the pair.
Neither of these plumes is strongly
detected in the GALEX or SDSS images,
although smoothed SDSS images show
a possible hint of the southern plume.
This is surprising in light of
the relatively high N$_{HI}$ in
the inner 5$'$ (10 kpc) sections of
these plumes of 4 $\times$ 10$^{20}$
cm$^{-2}$.
It was suggested by
\citet{clemens98a} that these HI features were produced
by SN-driven outflow from the main galaxy NGC 4490.
Instead, we suggest that
the HI plumes are simply gas-rich tidal features.
Deeper optical and UV imaging is needed to search for a stellar component
to these features.

\begin{figure}
\plottwo{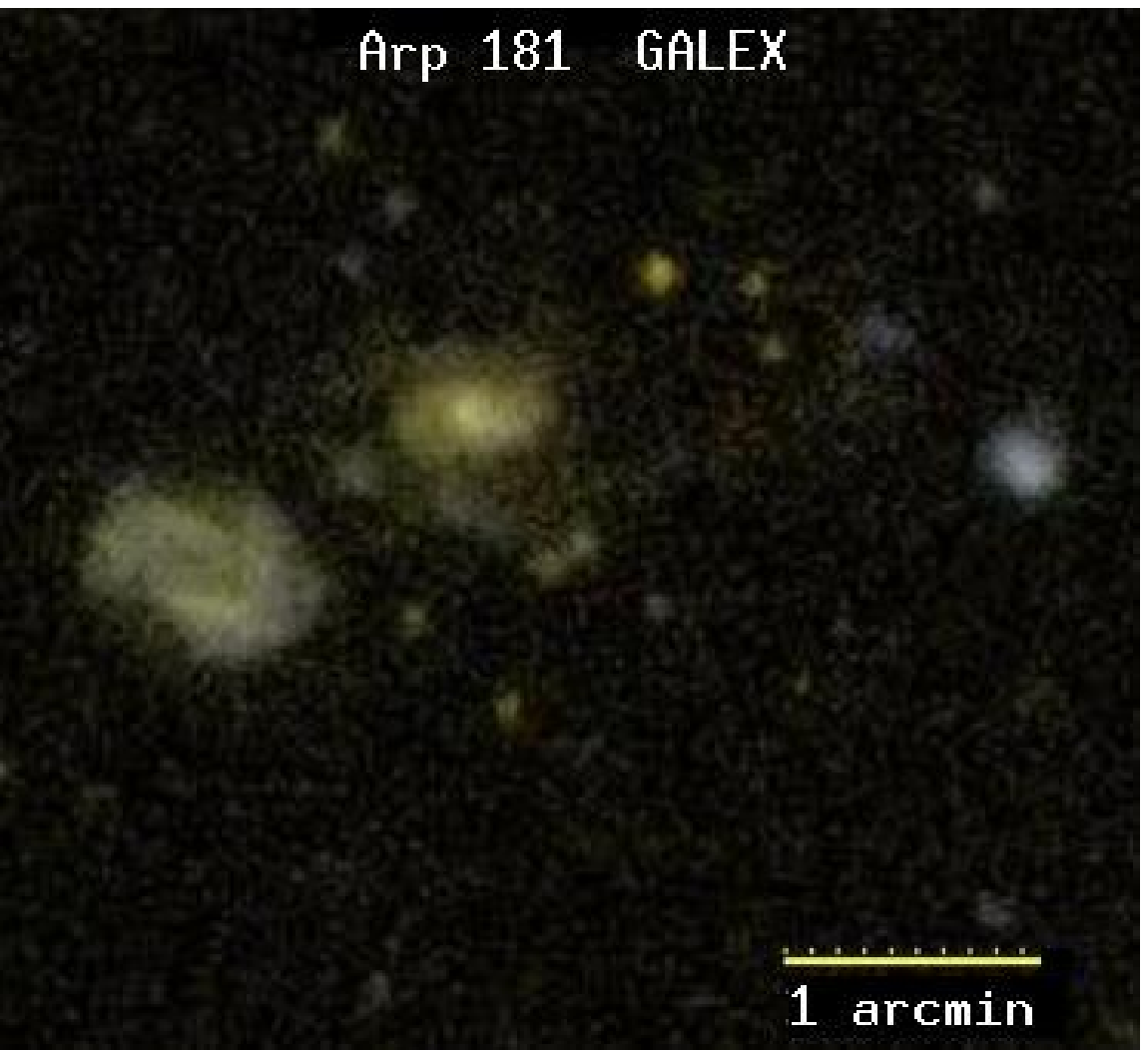}{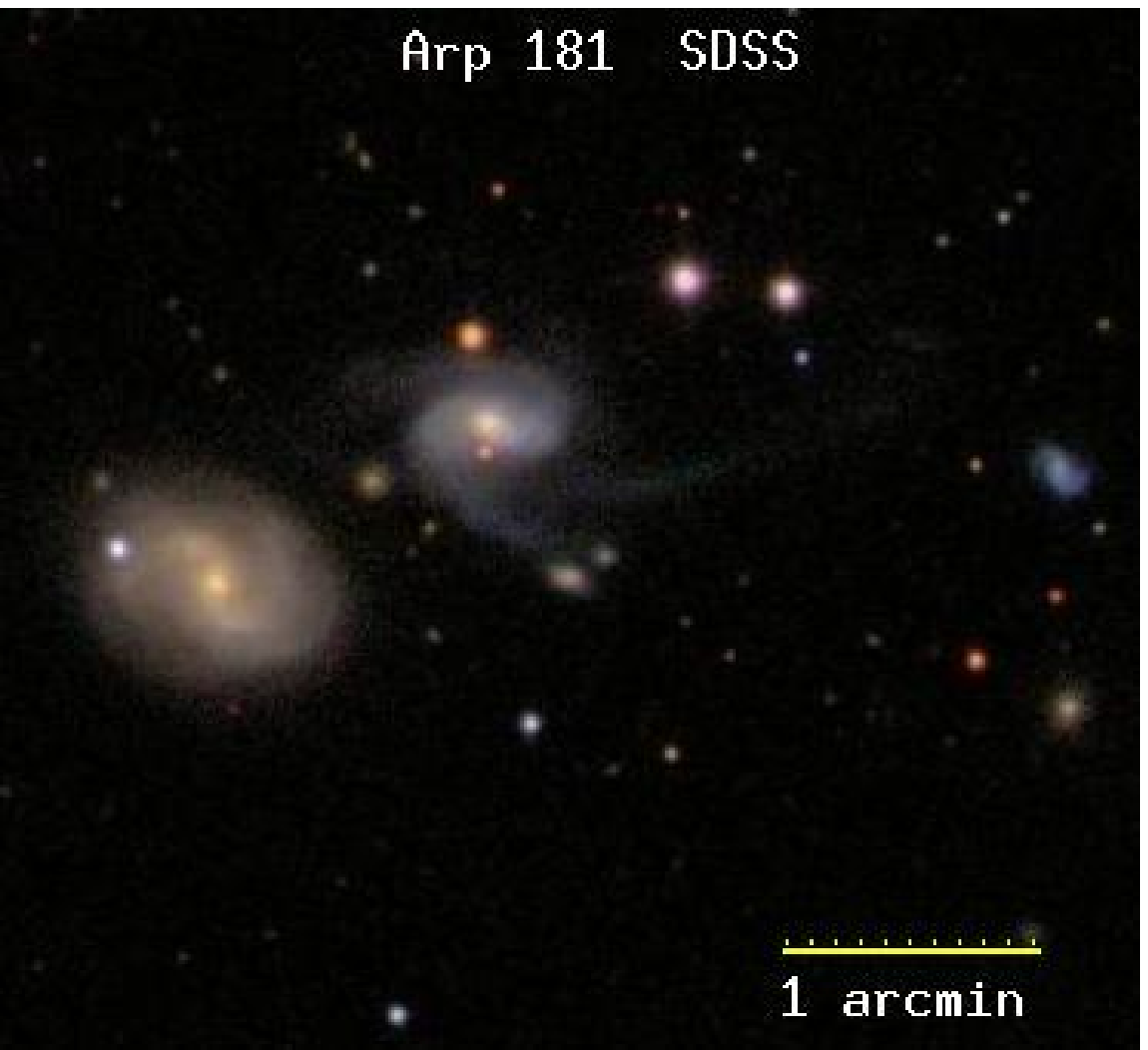}
\caption{
  \small 
The GALEX (left) and SDSS (right) images of Arp 181.
}
\end{figure}

\begin{figure}
\plottwo{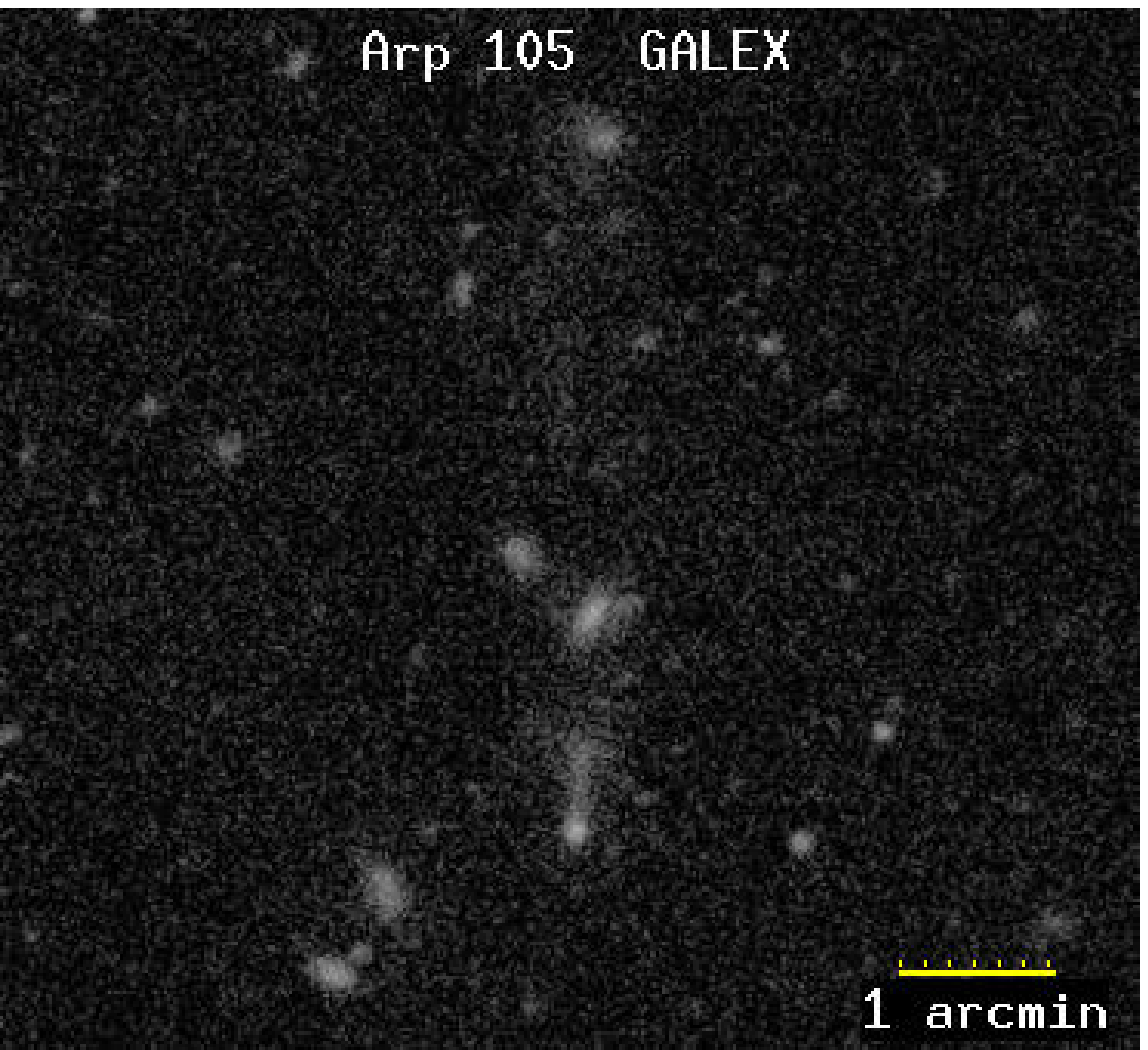}{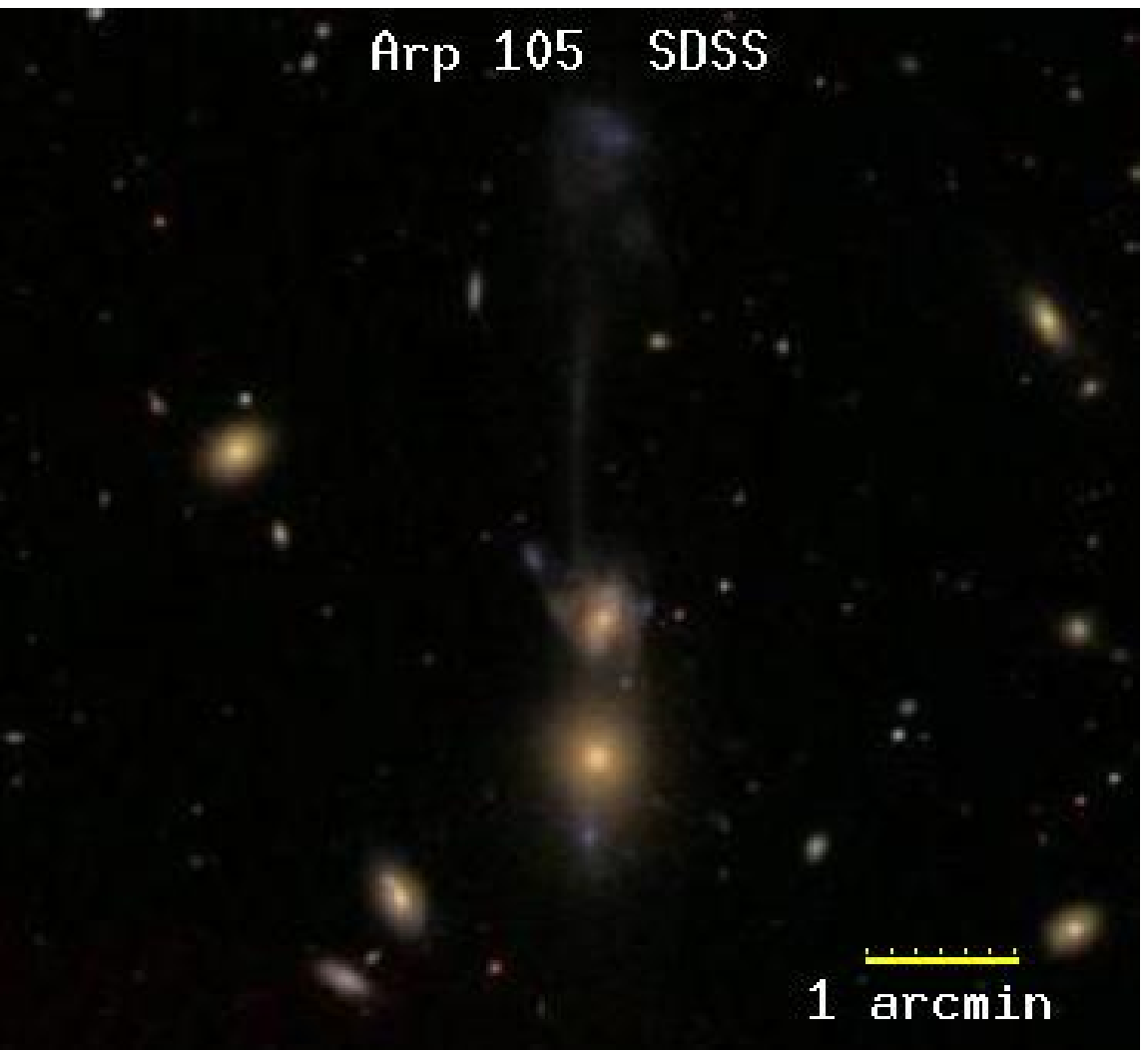}
\caption{
  \small 
The GALEX (left) and SDSS (right) images of Arp 105.
}
\end{figure}

\begin{figure}
\plottwo{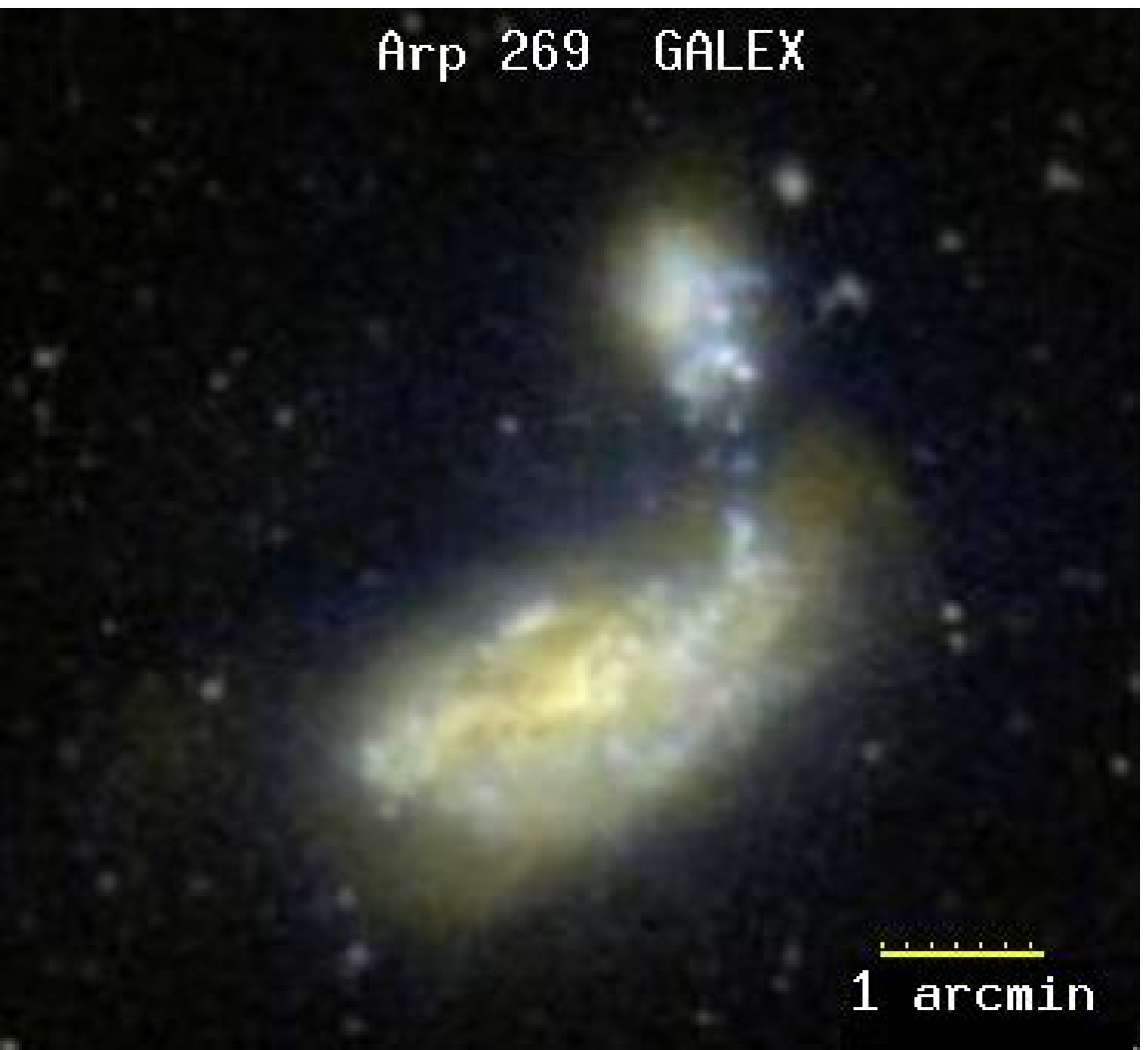}{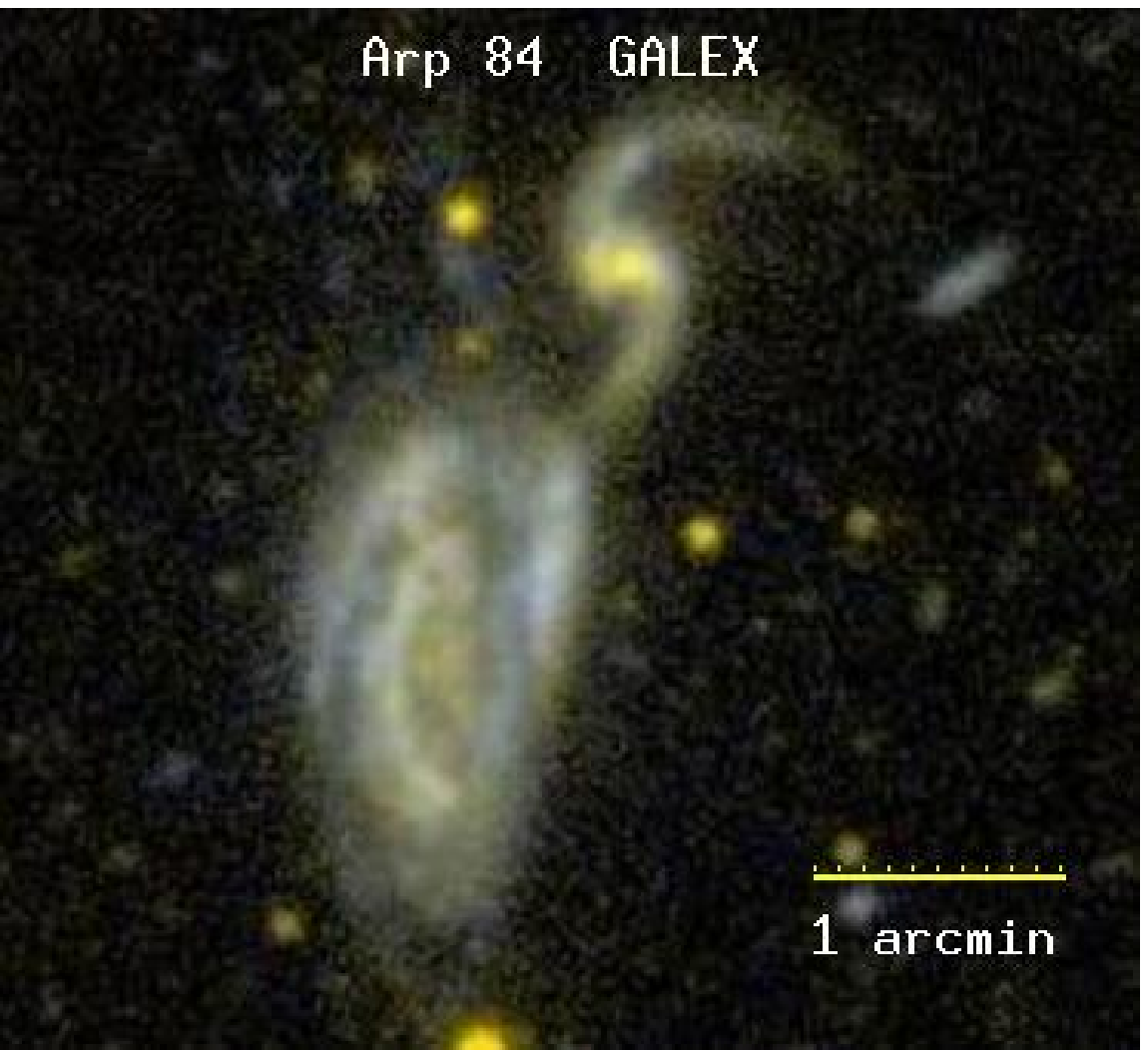}
\caption{
  \small 
The GALEX images of Arp 269 (left) and Arp 84 (right).
}
\end{figure}

Another system with a gas-rich tidal feature without
a GALEX counterpart is Arp 84
(`The Heron'),
a pair of unequal mass spiral galaxies
connected by a bridge (Figure 5b).
In HI maps, 
a large gaseous
plume extends to the 
south of the main galaxy 
\citep{kaufman99}.
In spite of
its 
high
N$_{HI}$ $\sim$ 4 $\times$ 10$^{20}$ cm$^{-2}$
and narrow HI line width of
$\sim$100 km~s$^{-1}$,
no diffuse UV emission is seen in this feature,
although
a few UV-bright clumps are present.
No redshifts are available at present
for these clumps.
Arp 84 also exhibits `beads on a string' 
along the inner edge of the northern tail of the smaller galaxy.
The most northern knot along this `string' is bright in
the GALEX image, as well as at 8 $\mu$m 
\citep{smith07} and H$\alpha$ \citep{kaufman99}.

\section{Summary}

In this paper, we present GALEX UV images of a subset
of the SB\&T interacting galaxy
pairs.
We identify examples of `beads on a string', accretion
from one galaxy to another, and candidate tidal dwarf 
galaxies.  
There are only a few tidal features that are bright in HI maps
that are not detected in GALEX.

\acknowledgements 

This research was supported by GALEX grant \\
GALEXGI04-0000-0026,
NASA LTSA grant NAG5-13079, and Spitzer
grant RSA 1353814.
We thank the Lick Observatory staff for their support of our optical
spectroscopy campaign.

\end{document}